\documentclass[pre,superscriptaddress,tightenlines,twocolumn,showpacs]{revtex4-1}

\usepackage{latexsym,amsmath,amsfonts,tabularx,amssymb,bm,empheq}
\usepackage{subfigure}
\usepackage{float}

\usepackage[normalem]{ ulem }
\usepackage{soul}

\usepackage{color}
\usepackage[dvipsnames]{xcolor}
\definecolor{nblue}{RGB}{28,130,185}

\definecolor{cgreen}{RGB}{76,153,0}

\definecolor{myorange}{RGB}{245,156,74}

\usepackage{hyperref}
\hypersetup{
  colorlinks=true,
  citecolor=magenta,
  urlcolor=-myorange
}

\newcommand{\bea}{\begin{eqnarray}}
\newcommand{\eea}{\end{eqnarray}}

\def\simge{\mathrel{%
   \rlap{\raise 0.511ex \hbox{$>$}}{\lower 0.511ex \hbox{$\sim$}}}}
\def\simle{\mathrel{
   \rlap{\raise 0.511ex \hbox{$<$}}{\lower 0.511ex \hbox{$\sim$}}}}

\def\simle{\mathrel{
   \rlap{\raise 0.511ex \hbox{$<$}}{\lower 0.511ex \hbox{$\sim$}}}}

\def\simge{\mathrel{%
    \rlap{\raise 0.511ex \hbox{$>$}}{\lower 0.511ex \hbox{$\sim$}}}}

\begin{document}


\title{A fixed point can hide another one: the nonperturbative behavior of \\ the tetracritical fixed point of the O($N$) models at large $N$}

\author{Shunsuke Yabunaka}\email{yabunaka123@gmail.com}
\affiliation{Advanced Science Research Center, Japan Atomic Energy Agency, Tokai, 319-1195, Japan}

\author{ Bertrand Delamotte}
\affiliation{
Sorbonne Universit\'e, CNRS, Laboratoire de Physique Th\'eorique de la Mati\`ere Condens\'ee, LPTMC, F-75005 Paris, France.
}

\date{\today}

\begin{abstract}

We show that at $N=\infty$ and below its upper critical dimension, $d<d_{\rm up}$, the critical and tetracritical behaviors of the O($N$) models are associated with the same renormalization group fixed point (FP) potential. Only their derivatives make them different with the subtleties that taking their $N\to\infty$ limit and deriving them do not commute and that two relevant eigenperturbations show singularities. This invalidates both the $\epsilon-$ and the $1/N-$ expansions. We also show how the Bardeen-Moshe-Bander line of tetracritical FPs at $N=\infty$ and $d=d_{\rm up}$ can be understood from a finite-$N$ analysis.
\end{abstract}

\maketitle

Field theories sometimes exhibit nonperturbative features such as confinement \cite{Pelaez21}, presence of bound states \cite{Hoyer21} or exotic excitations \cite{Guo19},  fixed points (FPs) of the renormalization group (RG) flows that are nonperturbative as in the Kardar-Parisi-Zhang equation \cite{Canet11}, divergence of the perturbative RG flow at a  finite RG scale \cite{Gredat}, presence of a cusp in the FP potential as in the random field Ising model \cite{tissier08}, to cite but a few. Very often, these nonperturbative effects  are assumed either  to occur in rather complicated theories such as gauge and string theories or in highly nontrivial statistical models. 

O($N$) models, which are the simplest scalar field theories, are often implicitly considered to be immune to these complex phenomena. Perturbative methods are therefore assumed to work almost all the time for these models, the exception to the rule being the Bardeen-Moshe-Bander (BMB) phenomenon \cite{Bardeen}, related to the existence of a line of tricritical FPs at $N=\infty$ and $d=3$, which requires nonperturbative FPs to be fully understood from a large-$N$ analysis \cite{Fleming20}. From this viewpoint, the enormous success of the $\epsilon=4-d$ expansion for the perturbative calculation of the critical exponents associated with the Wilson-Fisher (WF) FP \cite{Zinn-Justin} could let us believe that the critical physics of the O($N$) models is fully understood for any $N$ and  $d$, especially since it is corroborated by the $1/N$ and $\epsilon=d-2$ expansions \cite{Zinn-Justin}. 

Our goal in this Letter is to show instead that although the critical physics of the O($N$) models, described by the WF FP, is fully under perturbative control at both finite and infinite $N$, the tetracritical physics of these models at $N=\infty$ --and probably of infinitely many multicritical behaviors-- is not. We show below (i) that at $N=\infty$, it is also associated with the WF FP, which is unexpected, and (ii) that it nonetheless shows non-perturbative features that are beyond the reach of the standard implementation of both the large-$N$ and $\epsilon$- expansions. We show in particular a very intriguing phenomenon related to the large-$N$ limit of the tetracritical FP of the O($N$) models:  from the second order, the  derivatives  of the $N=\infty$ tetracritical FP potential, that is, of the WF FP potential, are not identical to the limit of the derivatives of the finite-$N$ tetracritical FP potentials when $N\to\infty$. This turns out to be crucial for understanding  the large-$N$ limit of tetracritical phenomena and shows that this limit is much less trivial than what is usually said \cite{BrezinWallace,Eyal,Zinn-Justin}.

The perturbative tetracritical FP corresponds to the massless $(\boldsymbol\varphi^2)^4$ theory, the upper critical dimension of which is $d_{\rm{up}}=8/3$.  It is found in perturbation theory in $\epsilon=8/3-d$ for all $N\ge1$ and it is three times infrared unstable \cite{Itzykson}. Calling $\lambda/(384 N^3)$ the coupling in front of the dimensionless $(\boldsymbol\varphi^2)^4$ term, the large-$N$ perturbative flow equation for $\lambda$ reads \cite{defenu2020}:
\begin{equation}
 \partial_t \lambda=-3\epsilon \lambda + \frac{9\lambda^2}{4N} + O(N^{-2}).
 \label{perturbation-theory-1}
\end{equation}
 From Eq. \eqref{perturbation-theory-1}, we find  that at leading order in $N$, the nontrivial FP solution is $\lambda^*=4\epsilon N/3$ from which follows that perturbation theory does not allow for a control of the large-$N$ limit of the tetracritical FP at fixed $\epsilon$. Only the double limit $N\to\infty$ and $\epsilon\to0$ such that the product $\epsilon N$ remains finite can possibly be under control. We come back on this point in the following.

Let us recall that in generic dimensions $d<4$, the only nontrivial FP found in the standard large-$N$ analysis {of the O($N$) models} is the WF FP \cite{Comellas}. Thus, no tetracritical FP is found  at $N=\infty$ and $d<8/3$ which is paradoxical considering that  it is perturbatively found for all  $N<\infty$ and $\epsilon>0$.

We show below that the solution to the paradox above lies in the field dependence of the tetracritical FP potential whereas it cannot be obtained from its field expansion and in particular  from $\lambda^*$. The  recourse to functional RG methods is therefore mandatory. 

The best way to implement functional RG is to consider Wilson's RG, as it is inherently functional \cite{Wilson}. We recall below the take-away philosophy of the modern version of Wilson's RG known as the nonperturbative  -- or functional -- renormalization group (NPRG).

 NPRG is based on the idea of integrating
fluctuations step by step \cite{PhysRevB.4.3174}. It is implemented on the Gibbs free energy $\Gamma$ \cite{wetterich91,wetterich93b,Ellwanger,Morris94,Berges,Delamotte-review,Delamotte-lect-notes} of a model defined by an Hamiltonian (or euclidean action) $H$ and a partition function ${\cal Z}$. To this model is associated a one-parameter family of models with Hamiltonians $H_k=H+ \Delta H_k$ and partition functions ${\cal Z}_k$, where  $k$ is a momentum scale. In $H_k$, $\Delta H_k$ is chosen such that only the rapid fluctuations in the original model, those with wavenumbers $\vert q\vert > k$, are summed over in the partition
function ${\cal Z}_k$. Thus, the slow modes ($\vert q\vert < k$) need to be decoupled in ${\cal Z }_k$ and this is achieved by giving them a mass of order $k$, that is by taking for $\Delta H_k$ a quadratic (mass-like) term, which is nonvanishing only for the slow modes:
\begin{equation}
 {\cal Z}_k[\boldsymbol{J}]= \int D\boldsymbol\varphi_i \exp(-H[\boldsymbol\varphi]-\Delta H_k[\boldsymbol\varphi]+ \boldsymbol{J}\cdot\boldsymbol\varphi)
 \label{partition-1}
\end{equation}
with $\Delta H_k[\boldsymbol\varphi]=\frac{1}{2}\int_q R_k(q^2) \varphi_i(q)\varphi_i(-q)$, where, for instance, $R_k(q^2)=(k^2-q^2) \theta(k^2-q^2)$ and $\boldsymbol{J}\cdot\boldsymbol\varphi=\int_x J_i(x) \varphi_i(x)$.
 The $k$-dependent  Gibbs free energy $\Gamma_k[\boldsymbol\phi]$
is defined as  the (slightly modified) Legendre transform of $\log  {\cal Z}_k[\boldsymbol{J}]$:
\begin{equation}
\label{legendre-1}
 \Gamma_k[\boldsymbol\phi]+\log  {\cal Z}_k[\boldsymbol{J}]= \boldsymbol{J}\cdot\boldsymbol\phi-\frac 1 2 \int_q R_k(q^2) \phi_i(q)\phi_i(-q)
 \end{equation}
with $\int_q=\int d^dq/(2\pi)^d$. With the choice of regulator function $R_k$ above, $\Gamma_k[\phi]$ interpolates between the Hamiltonian $H$ when $k$ is of order of the ultraviolet cut-off $\Lambda$ of the theory:  $\Gamma_\Lambda\sim H$,  and the Gibbs free energy $\Gamma$ of the original model when $k=0$:  $\Gamma_{k=0}=\Gamma$. 
The exact RG flow equation of $\Gamma_k$ gives the evolution of $\Gamma_k$ with $k$ between these two limiting cases. It is known as the Wetterich equation. It reads \cite{wetterich93b}:
\begin{equation}
\label{flow-1}
\partial_t\Gamma_k[\boldsymbol\phi]=\frac 1 2 {\rm Tr} [\partial_t R_k(q^2) (\Gamma_k^{(2)}[q,-q;\boldsymbol\phi]+R_k(q))^{-1}],
\end{equation}
where $t=\log(k/\Lambda)$, ${\rm Tr}$ stands for an integral over $q$ and a trace over group indices and $\Gamma_k^{(2)}[q,-q;\boldsymbol\phi]$ is
the matrix of the Fourier transforms of $\delta^2\Gamma_k/\delta \phi_i(x)\delta \phi_j(y)$. 

In most cases, Eq. (\ref{flow-1}) cannot be solved exactly and  approximations are mandatory. The best known approximation consists in expanding $\Gamma_k$ in powers of  the derivatives of $\phi_i$ and to truncate the expansion at a given finite order\cite{canet03,canet05,kloss14,delamotte04,benitez08,canet04,canet16,leonard15,balog19}. The approximation at lowest order is dubbed the local potential approximation (LPA). For the O($N$) model it consists in approximating  $\Gamma_k$ by:
\begin{equation}
 \Gamma_k[\boldsymbol\phi] = \int_{x} \left(\frac{1}{2} (\nabla \phi_i)^2 + U_k(\phi)\right)     
   \label{ansatz-order2-1}
\end{equation}
where $ \phi=\sqrt{\phi_i\phi_i}$. Fixed points are found only for dimensionless quantities and the standard large-$N$ limit  by rescaling the field and the potential by  factors $N^{-1/2}$ and $N^{-1}$ respectively. Thus, we define the dimensionless and rescaled field $\bar{\phi}$ and potential $\bar{U}_{k}$
as 
$\bar{\phi}=v_{d}^{-\frac{1}{2}}k^{\frac{2-d}{2}}N^{-1/2}\phi$ and
$\bar{U}_{k}(\bar{\phi})=v_{d}^{-1}k^{-d}N^{-1}U_{k}\left(\phi\right)$
with 
$v_{d}^{-1}=2^{d-1}d\pi^{d/2}\Gamma(\frac{d}{2})$. 
The LPA flow of $\bar{U}_{k}$ then reads:
\begin{equation}
   \begin{split}
 \partial_t\bar{U}_k(\bar\phi)=&-d\,\bar U_{k}(\bar\phi)+\frac{1}{2}(d-2)\bar\phi\, \bar U_{k}'(\bar\phi)+\\  
&\left(1-\frac{1}{N}\right)\frac{\bar\phi}{\bar\phi+\bar U_{k}'(\bar\phi)}+\frac{1}{N}\frac{1}{1+\bar U_{k}''(\bar\phi)}\hspace{0.5cm}
\end{split}
\label{flow-LPA-Ubar-1}
\end{equation}
with $\partial_t=k\partial_k$.
The standard large-$N$ limit of the LPA flow equation above  is obtained by (i) replacing the factor $1-1/N$ by 1, (ii) dropping the last term   in Eq. (\ref{flow-LPA-Ubar-1}) because it is assumed to be sub-leading \cite{Tetradis-Litim}.
As a consequence of the two steps above, the explicit dependence in $N$ in Eq. \eqref{flow-LPA-Ubar-1} disappears in the large-$N$ limit. 

The crucial point of the large-$N$ limit is that assuming point (ii) above, the resulting LPA flow equation on $\bar U_k$ can be shown to be  {\it exact} in the limit $N\to\infty$ \cite{dattanasio}. Under this assumption, all FPs of the O($N)$ models have been found exactly at $N=\infty$ \cite{Tetradis-Litim,Comellas,Kubyshin,dattanasio,Katis-Tetradis}. The result is the following: In a generic dimension $d<4$  there is only one nongaussian FP at $N=\infty$ which is the usual Wilson-Fisher FP (WF). The exceptions to the rule above are the BMB lines of FPs \cite{Bardeen,Comellas,David,Omid,Mati2017} existing in dimensions $d=2+2/p$ with $p$ an integer larger than 1.

We now show that the procedure described above  is too restrictive to study the large-$N$ limit of the tetracritical FPs. As said above, the standard large-$N$ analysis consists in neglecting  the last term in Eq. (\ref{flow-LPA-Ubar-1}). However, this term is negligible  only if $(1+\bar U_{k}''(\bar\phi))^{-1}$ does not counterbalance at large $N$ its $1/N$ prefactor for some finite values of $\bar\phi$.  We now show that  because of singularities in the third derivative of $\bar U_{k}(\bar\phi)$, the contribution of the last term in Eq. (\ref{flow-LPA-Ubar-1}) cannot be neglected in the FP equation of $\bar U_{k}''(\bar\phi)$ obtained by differentiating twice Eq.~\eqref{flow-LPA-Ubar-1} (see footnote below Eq. (\ref{eq:ON-1-1}) for more detail). This turns out to be sufficient to invalidate the standard large-$N$ limit in the tetracritical case.


\begin{figure}
\includegraphics[scale=0.28]{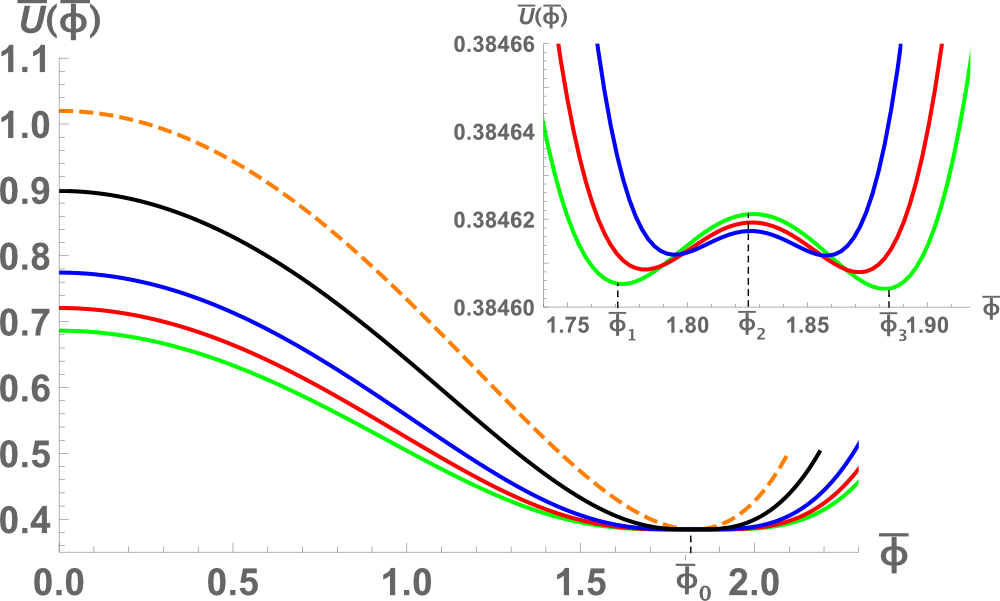}
 \caption{$d=2.6$: $\bar{U}(\bar{\phi})$ for the $T_{3}$ FP  of Eq. (\ref{flow-LPA-Ubar-1}). Green, red, blue and black curves correspond to $N=1500$, 2250, 4500 and 42000. The orange dashed curve corresponds to the WF FP at $N=\infty$. Inset: Close view of $\bar{U}(\bar{\phi})$ around $\bar\phi_i$. }
\label{T3FP2-1}
\end{figure}

We have numerically solved Eq. (\ref{flow-LPA-Ubar-1}) and have found for several values of $N$ and $d<8/3$ the perturbative tetracritical FP that we call $T_3(N,d)$. As expected,  $T_3$ bifurcates from the Gaussian FP in $d=8/3^-$. We have followed it  down to $d=2.6$, see Fig. \ref{T3FP2-1}  and Fig. \ref{T3FP1-1} of the Suppl. Mat. The FP potential of $T_3$, (i) shows as expected two maxima, one of which being located at $\bar\phi=0$ and another one at $\bar\phi_2>0$, and two minima at $\bar\phi_1$ and $\bar\phi_3$ such that $\bar\phi_3>\bar\phi_2>\bar\phi_1>0$, see Fig.  \ref{T3FP2-1},  (ii)  can be continuously followed up to arbitrarily large values of $N$ at fixed  $d<8/3$, (iii) has its three extrema $\bar\phi_1, \bar\phi_2, \bar\phi_3$  approaching each other when $N$ is increased at fixed $d$. These extrema  tend to a common value $\bar\phi_0$ when $N\to\infty$ which is the minimum of the FP potential, see Fig. \ref{T3FP2-1} and Fig. \ref{3extrema-loglog-1} of the Suppl. Mat. Point (ii) above is paradoxical because it seems to contradict the standard large-$N$ approach where only the WF FP is found in a generic dimension $d<8/3$ at $N=\infty$. We now show  that the  WF FP potential at $N=\infty$ is in fact the limit when $N\to\infty$ of the potential of $T_3$ for $d<8/3$. This solves the above paradox because it explains why on one hand there exists a nontrivial tetracritical FP at $N=\infty$ and $d<8/3$ and on  the other hand that there is no other nontrivial and smooth solution of Eq. \eqref{flow-LPA-Ubar-1} at $N=\infty$ than the WF FP potential. However, this creates a new paradox since obviously the critical and tetracritical universal behaviors cannot be the same since the two FPs do not have the same number of unstable eigendirections.  
We now explain in detail this new paradox. 

We can see on Fig. \ref{T3FP2-1}  that the FP potentials found in $d=2.6$ for large values of $N$ are extremely flat in the region,  $\bar\phi\in[\bar\phi_1,\bar\phi_3]$ because the three extrema  are very close  and the height of the barrier between the two minima very small. We have numerically found that the height of the barrier scales as $N^{-1}$ and the distance between the two minima  as $N^{-1/2}$ so that the curvatures $\bar U'' (\bar\phi_i)$ at the three extrema approach constant values as $N\to\infty$, see Fig. \ref{3extrema-loglog-1} of the Suppl. Mat. This suggests that $\bar U'' (\bar\phi)$ while being well-behaved everywhere but between the three extrema, changes very rapidly within a boundary layer around  $\bar\phi_0$ of typical width $N^{-1/2}$, making divergent $\bar U''' (\bar\phi_0)$ when $N\to\infty$.  

It is not common in physics to encounter this kind of situation where a series of functions $f_n(x)$ tends to a smooth function $f_\infty(x)$ whereas from a certain order $p$, their derivatives $f_n^{(p)}(x)$ do not tend to $f_\infty^{(p)}(x)$. However, a simple toy model explains trivially how this can occur. Consider the series of functions $f_n(x)=n^{-1}\sin(n^2 x)$. Obviously, $f_\infty(x)\equiv0$ which implies that $f_\infty'(x)\equiv0$ whereas $\lim_{n\to\infty}f_n'(0)=\infty$. 
 
  In our case, at fixed $d<8/3$, the limit of the $T_3$ potentials when $N\to\infty$ is a nontrivial and well-defined function that therefore must be the WF FP potential. {We have checked that  it is indeed the limit of $T_3$ when $N\to\infty$, see Fig. \ref{T3FP2-1}}. The difference between the critical and tetracritical behaviors is therefore not visible on the  potentials themselves but only on their derivatives as we now show. 
  
Let us study the boundary layer around $\bar\phi_0$. It is convenient for what follows to change variables. Following Ref. \cite{Morris}, we define: $V(\mu)=U(\phi)+(\phi-\Phi)^2/2$ with $\mu=\Phi^2$ and $\phi-\Phi=-2\Phi V'(\mu)$. As above, it is convenient to rescale $\mu$ and $V(\mu)$:   $\bar\mu=\mu/N$, $\bar V=V/N$. In terms of these quantities, the FP equation for $\bar V(\bar\mu)$ reads 
\begin{equation}
 0=1-d\,\bar V+(d-2)\bar\mu \bar V'+4\bar\mu{\bar V'}{}^2-2\bar V'-\frac{4}{N}\bar\mu\,\bar{V}''.
\label{flow-LPA-WP-essai-1}
\end{equation}
Eq. \eqref{flow-LPA-WP-essai-1} has two remarkable features: (i) it is much simpler than Eq. (\ref{flow-LPA-Ubar-1}) because the nonlinearity comes only from the $(\bar V')^2$ term, (ii) it is the LPA equation obtained from the Wilson-Polchinski (WP) version of the NPRG \cite{Wilson, Polchinski, Hasenfratz}. Thus, $\bar V(\bar\mu)$  is related to the potential $\bar U(\bar\phi)$ of the Wetterich version of the RG by the Legendre transform of Eq. (\ref{legendre-1}). The standard large-$N$ analysis performed in this version of the NPRG consists here again in neglecting the last term in  Eq. (\ref{flow-LPA-WP-essai-1}) because it is suppressed by a  $1/N$ factor. Under the assumption that this term is indeed negligible, the resulting equation can be solved exactly in the large-$N$ limit \cite{Kubyshin,Comellas}. However, at large $N$, it is clear on Eq. (\ref{flow-LPA-WP-essai-1}) that we have to deal with singular perturbation theory since the small parameter used in the $1/N$ expansion is in front of the term of highest derivative, that is,  $\bar V''$. In this case, it is well-known that at large $N$ a boundary layer can exist for a particular value of  $\bar\mu$  that becomes a singularity at $N=\infty$, making this term non negligible \cite{Holmes}.

  The value of $\bar\mu$  corresponding to $\bar\phi_0$  is called $\bar\mu_0$ and is the minimum of $\bar V(\bar\mu)$ at $N=\infty$.  We find for $\bar V(\bar\mu)$ the same features about its three extrema $\bar\mu_i$ as for $\bar U(\bar\phi)$ at $\bar\phi_i$: The three extrema $\bar\mu_i$ approach each other and to $\bar\mu_0$ as $N\to\infty$,  the distances between them scale as $N^{-1/2}$ and the curvatures $\bar V'' (\bar\mu_i)$  as $N^{0}$. Taking into account the scaling around $\bar\mu_0$ inside the boundary layer, we introduce another scaled variable $\tilde\mu=N^{1/2}(\bar\mu-\bar\mu_0)$. 
Since at $N=\infty$, $\bar{V}'(\bar\mu)$ vanishes at $\bar\mu=\bar\mu_0$,  $\bar{V}(\bar\mu_0)$ should approach $1/d$ at  leading order in $N^{-1/2}$. We therefore define a scaled boundary layer by  $\tilde{V}_N(\tilde\mu)=N\left( \bar{V}(\bar\mu_0+N^{-1/2} \tilde\mu )-1/d\right)$
  which implies $\tilde{V}_N''(\tilde\mu)=\bar{V}''(\bar\mu_0+N^{-1/2}\tilde\mu)$. We plot $\tilde{V}_N''(\tilde\mu)$ for several values of $N$ in Fig. \ref{boundarylayer-1} of the Suppl. Mat. 
 
\begin{figure}
\includegraphics[scale=0.8]{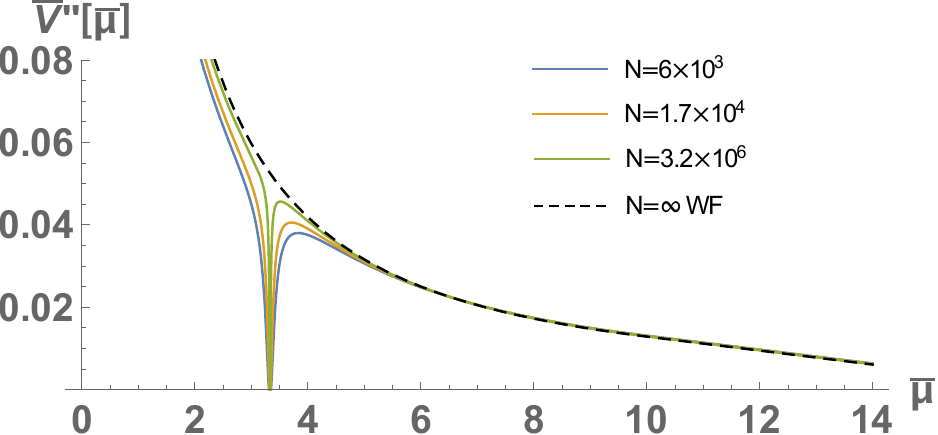}
 \caption{Second derivative of the WF and $T_3$ FP potentials for different values of $N$ in $d=2.6$. }
\label{fig:V''-1}
\end{figure}

By substituting $\tilde{V}_N(\tilde\mu)$ by its value in Eq. (\ref{flow-LPA-WP-essai-1}) and solving it at  order  $O(N^{-1/2})$, we find that $\bar\mu_0=2/(d-2)$. 
At  order  $O(N^{-1})$, Eq. (\ref{flow-LPA-WP-essai-1}) becomes
\begin{equation}
-\frac{8 \tilde{V}''_{\infty}(\tilde\mu)}{d-2}+\frac{8 \tilde{V}'_{\infty}(\tilde\mu)^2}{d-2}+(d-2) \tilde\mu\tilde{V}'_{\infty}(\tilde\mu)-d \tilde{V}_{\infty}(\tilde\mu)=0\label{eq:ON-1-1}
\end{equation}
 \cite{footnote1} which  is clearly invariant under $\tilde\mu\to -\tilde\mu$ from which it follows that  $\tilde{V}_{\infty}'(0)=0$. At  $\tilde\mu=\infty$,   $\tilde{V}''_{\infty}(\tilde\mu)$ should tend to a finite value that matches with $\bar{V}''(\mu)$ at $\bar\mu_0^+$. This implies that the solution of Eq. \eqref{eq:ON-1-1} should be quadratic when $\tilde\mu\to\infty$. Substituting $\tilde{V}_\infty(\tilde\mu)$ by $\tilde{V}''_{\infty}(\tilde\mu=\infty) \tilde\mu^2/2$ in Eq. (\ref{eq:ON-1-1}) and balancing the leading terms as $\tilde\mu\rightarrow \infty$, we find that  $\tilde{V}''_{\infty}(\tilde\mu=\infty)=(-d^2+6d-8)/16$.  Imposing the two boundary conditions found above at $\tilde\mu=0$ and $\tilde\mu=\infty$ selects a unique and globally defined solution $\tilde{V}''_{\infty}(\tilde\mu)$ of Eq. (\ref{eq:ON-1-1}) shown in Fig. \ref{boundarylayer-1} of the Suppl. Mat.  We find  $\bar{V}''(\bar{\mu}_0^+)=\bar{V}_{\rm WF}''(\bar{\mu}_0)=\tilde{V}''_{\infty}(\tilde\mu=\infty)$ which proves the matching at $N=\infty$ between the boundary layer and the potential outside of the layer, see Fig. \ref{fig:V''-1}. We have shown in Fig. \ref{fig:U''-1} of the Suppl. Mat. the boundary layer for $\bar U''(\bar\phi)$ analogous to that of $\bar{V}''(\bar{\mu})$.   To conclude, we have proven that for $d<8/3$, a boundary layer develops at  large $N$  for the second derivative of the $T_3$ potential {that becomes a singularity when $N\to\infty$}. What remains to be understood is its physical relevance.

At first sight, what we have obtained for $T_3$ looks paradoxical because we could think that its potential being identical to the WF potential at $N=\infty$,  the linearized flow around these two FPs should also be identical and thus the same for all critical exponents. We now show that this naive argument is wrong.

We have computed in $d<8/3$ the relevant eigenvalues of the RG flow around $T_3$ and WF at finite and large $N$ and as expected we have  found three for $T_3$ and one for WF. When $N\to\infty$, one of the three eigenvalues at $T_3$ tends as expected to $d-2$ which is the relevant eigenvalue $\nu^{-1}$ of the critical WF FP at $N=\infty$ \cite{Zinn-Justin,Comellas}. The nontrivial point is that the two other relevant eigenvalues at $T_3$ have a well-defined limit when  $N\to\infty$ although they do not play any role for the critical behavior of the O($N=\infty$) model. The solution to this paradox is that they are associated with eigenperturbations  that become singular when $N\to\infty$. That these two eigenperturbations become singular is clear for one of them, called $\delta \bar V_2$, on Fig. \ref{fig:eigenperturbation-1,2-1} of the Suppl. Mat. As for the other one, $\delta \bar V_1$,  its slope at $\bar\mu_0$ diverges as $N^{1/3}$ which implies that at $N=\infty$, it becomes discontinuous at $\bar\mu_0$, see Figs. \ref{fig:eigenperturbation-1,2-1} and \ref{fig:slope-1} of the Suppl. Mat.  For ordinary second order phase transitions, these eigenperturbations are excluded which explains that the associated relevant eigenvalues do not play any role. This solves all the paradoxes associated with the tetracritical FPs at $N=\infty$ and $d<8/3$.

What remains to be studied is the particular case $N=\infty$  and $d=8/3$ where a line, called the BMB line, of smooth tetracritical FPs shows up. It is obtained in the WP version of the RG by integrating Eq. \eqref{flow-LPA-WP-essai-1} in which the last term, proportional to $1/N$, has been discarded. It is given by the following implicit expression  \cite{Comellas}:
\begin{equation}
\bar{\mu}_{\pm}=\frac{ C}{\bar{V}'\left(1-2\bar{V}'\right)}\left(\frac{\pm2\bar{V}'}{1-2\bar{V}'}\right)^{4/3}+2 f(4\bar{V}'),\label{eq:BMB-line-1}\end{equation}
where $f(x)$, which is analytic for $x<2$, is given by 
\begin{equation}
f(x)=\frac{3}{2-x}+\frac{4x}{\left(2-x\right)^{7/3}}\int_{0}^{1}dz\left(\frac{2-xz}{z}\right)^{1/3}\label{eq:fvprime-1}\end{equation}
and $\bar{\mu}_{\pm}$ correspond to the two branches $\bar{\mu}>3$ and $\bar{\mu}<3$, respectively. The derivative of the potential  $\bar{V}'$ is positive (negative) on the former (latter) branch and $C$ is a non-negative integration constant.  $\bar{V}(\bar{\mu})$ is analytic at $\bar\mu=\bar\mu_0=3$ and  $\bar{V}'(\bar{\mu}=3)=0$. In Fig. \ref{fig:BMB-1} of the Suppl. Mat. different $\bar{V}'(\bar{\mu})$ corresponding to different FPs of the BMB line are shown.  All FPs along the BMB line share the same critical exponents, that is, the exponents of the Gaussian FP which is itself tetracritical. Notice that the WF FP which corresponds to $C=0$, is the  end point of this line and deserves special attention. We come back on this point in the following.

From Eq. \eqref{perturbation-theory-1}, we have seen  that $\lambda^*$ remains constant  at leading order in $1/N$ along the hyperbola of constant $\epsilon N $ of the $(d,N)$ plane. This suggests that when the double limit $d\to 8/3$ and $N\to\infty$ is taken at fixed $\alpha=\epsilon N $,  $T_3$  converges in $d=8/3$ to one of the FPs of the BMB line. We have analytically and numerically checked  this and have derived analytically the relation between $\alpha$ and $C$: $\alpha=162/C^3$, see Suppl. Mat. and Fig. \ref{fig:fixedalpha-1}.


Two extreme cases are worth studying. First, the Gaussian FP  corresponds to the limit $N\to\infty$ at fixed dimension $d=8/3$, that is, at $\alpha=0$. It corresponds to  $C=\infty$ in Eq. \eqref{eq:BMB-line-1}. Second, $\alpha=\infty$, which implies $C=0$, corresponds to taking the limit $\epsilon\to 0$ at fixed $N=\infty$, that is, to following the WF FP at $N=\infty$ up to $d=8/3$. However, at finite $\epsilon$ and $N=\infty$, we know from the analysis above that the last term in Eq. \eqref{flow-LPA-WP-essai-1}  cannot be neglected. Consistently,  the same occurs for the BMB line: the WF FP potential is indeed the end point of the BMB line obtained by taking the limit $C\to0$ in Eq. \eqref{eq:BMB-line-1} but the derivatives of this potential can only be studied by retaining the last term in Eq. \eqref{flow-LPA-WP-essai-1}.  Here again, this explains why the $T_3$ FP in the $C\to 0$ limit  is three times unstable and not only once unstable. 

To conclude, we have solved the paradox of the apparent absence of a nontrivial tetracritical FP at $N=\infty$ and $d<8/3$ by showing that this FP does exist but is nothing else than the WF FP up to the subtlety that the derivatives of the tetracritical FP potential are not the derivatives of the WF FP potential. This makes the large-$N$ limit of the O($N$) model much less trivial than is usually advocated at least for multicritical phenomena. The fact that the tetracritical FP has two more unstable infrared directions than the WF FP is related to this subtle point because they are associated with singular eigenperturbations, a possibility which is usually not considered. We conjecture that what  has been found above at large $N$ and for $d\le8/3$  is valid for all multicritical points with an odd number of eigendirections below or at their upper critical dimension because the BMB lines for all of them terminate at the WF FP \cite{Comellas}, a fact that in itself is almost enough to imply everything else.
Let us finally point out that what we have found for the tetracritical FP is very different from what was found around $d=3$ at large-$N$ in the tricritical case which required the existence of new FPs to be fully understood at finite $N$ \cite{Pisarski,Osborn,Yabunaka-Delamotte,Yabunaka-PRE2022}. We also conjecture that this phenomenon is not specific to the O($N$) models but should rather be generic.

We acknowledge A. Codello and N. Defenu  and  for correspondence and discussions at an early stage of this work.  S. Y. was supported by Grant-in-Aid for Young Scientists  (18K13516).

\newpage

\centerline{\bf SUPPLEMENTAL MATERIALS}

\section{$T_3$ FP potentials in $d<8/3$}
\begin{figure}[H]
\includegraphics[scale=0.5]{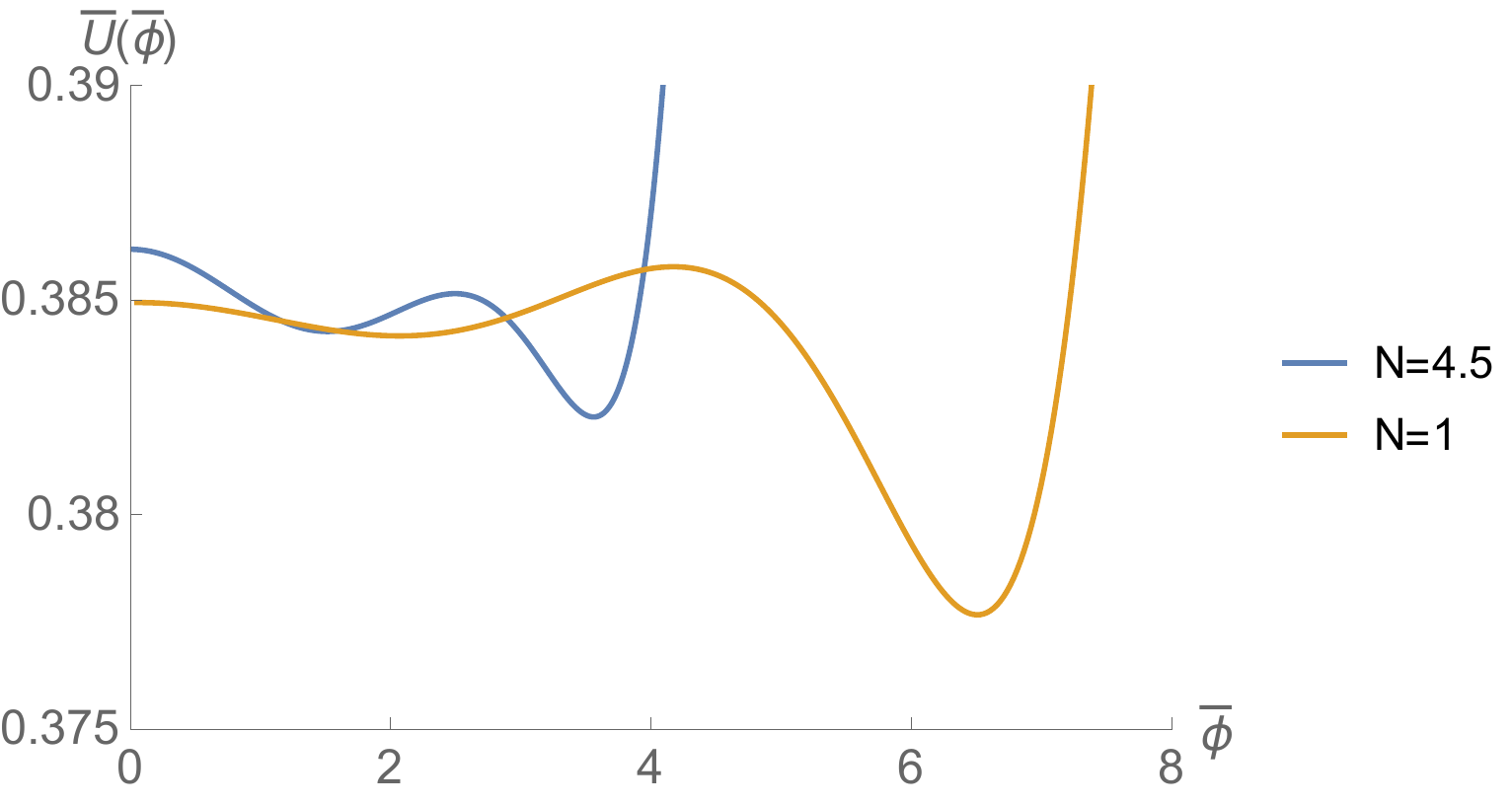}
 \caption{$\bar{U}(\bar{\phi})$ for the $T_{3}$ FP  for different values of $N$ in $d=2.6$. }
\label{T3FP1-1}
\end{figure}

We show in Fig. \ref{T3FP1-1} the tetracritical FP potential $\bar U(\bar\phi)$  obtained with the LPA and solution of  Eq. \eqref{flow-LPA-Ubar-1} for  small values of $N$. They have the typical shape of a tetracritical potential showing two nontrivial minima.

\section{Large-$N$ behavior of the extrema of the tetracritical potential}
\begin{figure}[H]
\includegraphics[scale=0.4]{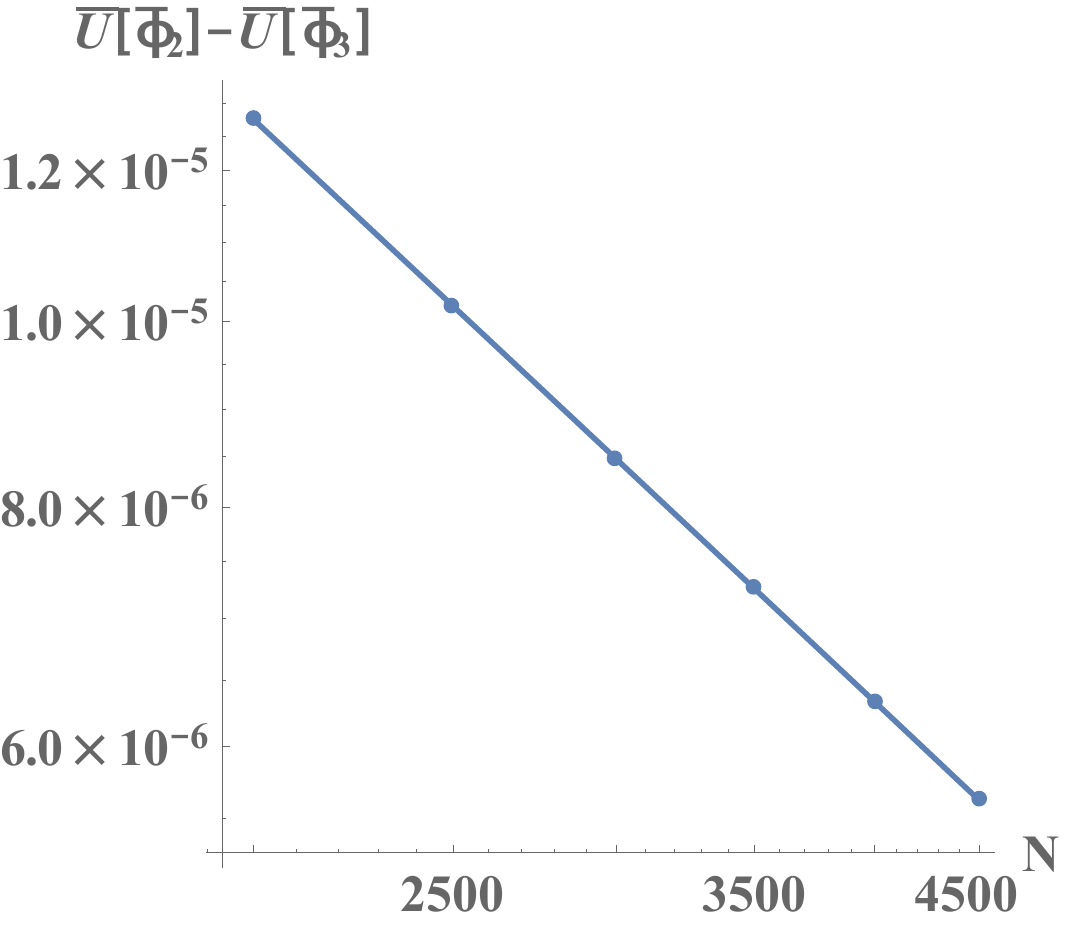}
\includegraphics[scale=0.4]{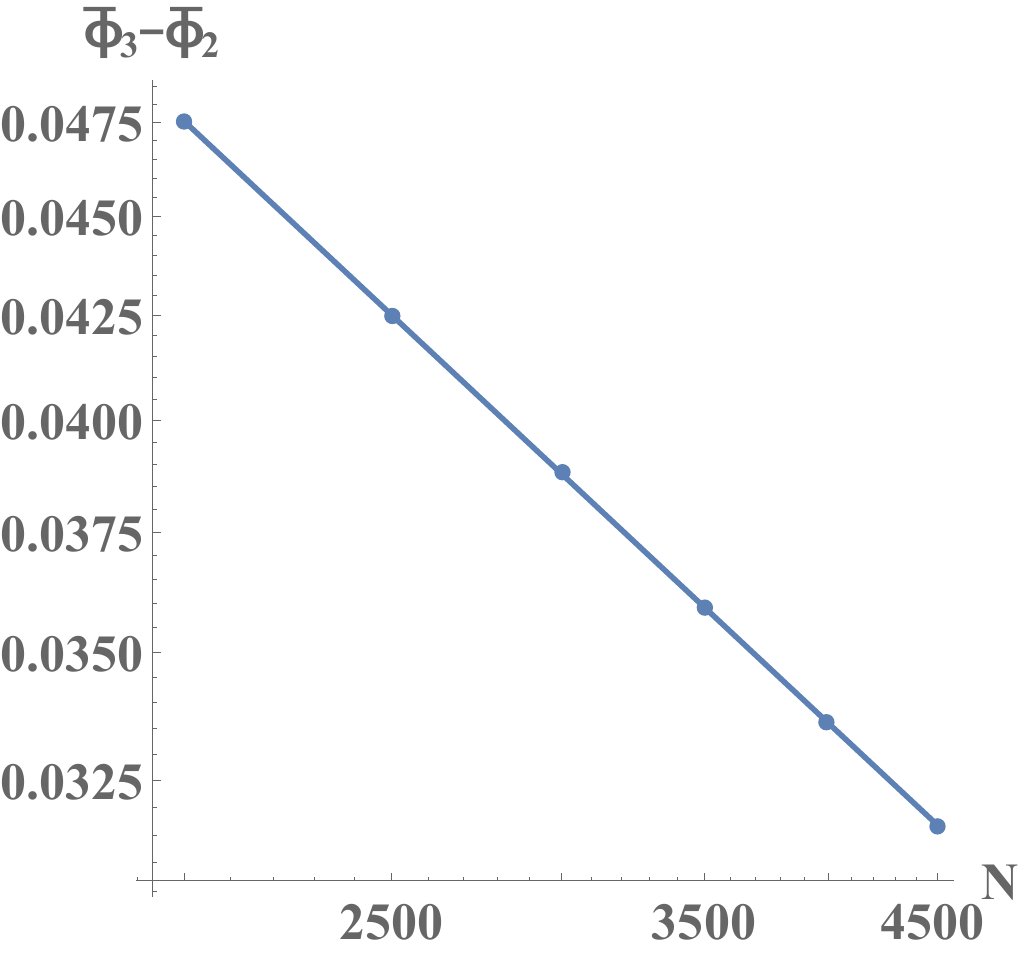}
 \caption{Left: Height of the potential barrier for the $T_{3}$ FP of Eq. (\ref{flow-LPA-Ubar-1}) for large values of $N$ in $d=2.6$ (blue dots). The equation of the full line is  $y=0.0257/N$. Right: Distance between the maximum $\bar{\phi}_2$ and the minimum $\bar{\phi}_3$   for the $T_{3}$ FP of Eq. (\ref{flow-LPA-Ubar-1}) for large values of $N$ in $d=2.6$ (blue dots). The equation of the full line is $y=2.12506/N^{1/2}$. }
\label{3extrema-loglog-1}
\end{figure}

The three nontrivial extrema of the $T_3$  FP potential in either the WP or Wetterich version of the RG, shown in Fig. \ref{T3FP2-1} of the main text,  behave the same way when $N\to\infty$. We show on Fig. \ref{3extrema-loglog-1} the scaling in $N$ of the height of the barrier between the extrema $\bar\phi_i$ of the rescaled potential $\bar U(\bar\phi)$ of the Wetterich version of the RG, as well as the distance between them. These extrema are shown in Fig. \ref{T3FP2-1} of the main text.
Since the height of the barrier, $\Delta \bar U$, scales as $N^{-1}$ and the distance between the extrema, $\Delta \bar \phi$, as $N^{-1/2}$, a simple dimensional argument shows that the curvatures at these extrema that goes as  $\Delta \bar U/(\Delta \bar\phi)^2$, do not scale with $N$, that is, tend to constants when $N\to\infty$, a fact that we have numerically checked. Thus, for $d<8/3$ and at large and finite $N$, the curvature of $\bar U(\bar\phi)$ varies between a positive value at $\bar\phi_1$, a negative value at $\bar\phi_2$ and again a positive value at $\bar\phi_3$ on a distance of order $N^{-1/2}$.

\section{The scaled boundary layer  $\tilde V''(\tilde\mu)$}
\begin{figure}[H]
\includegraphics[scale=0.5]{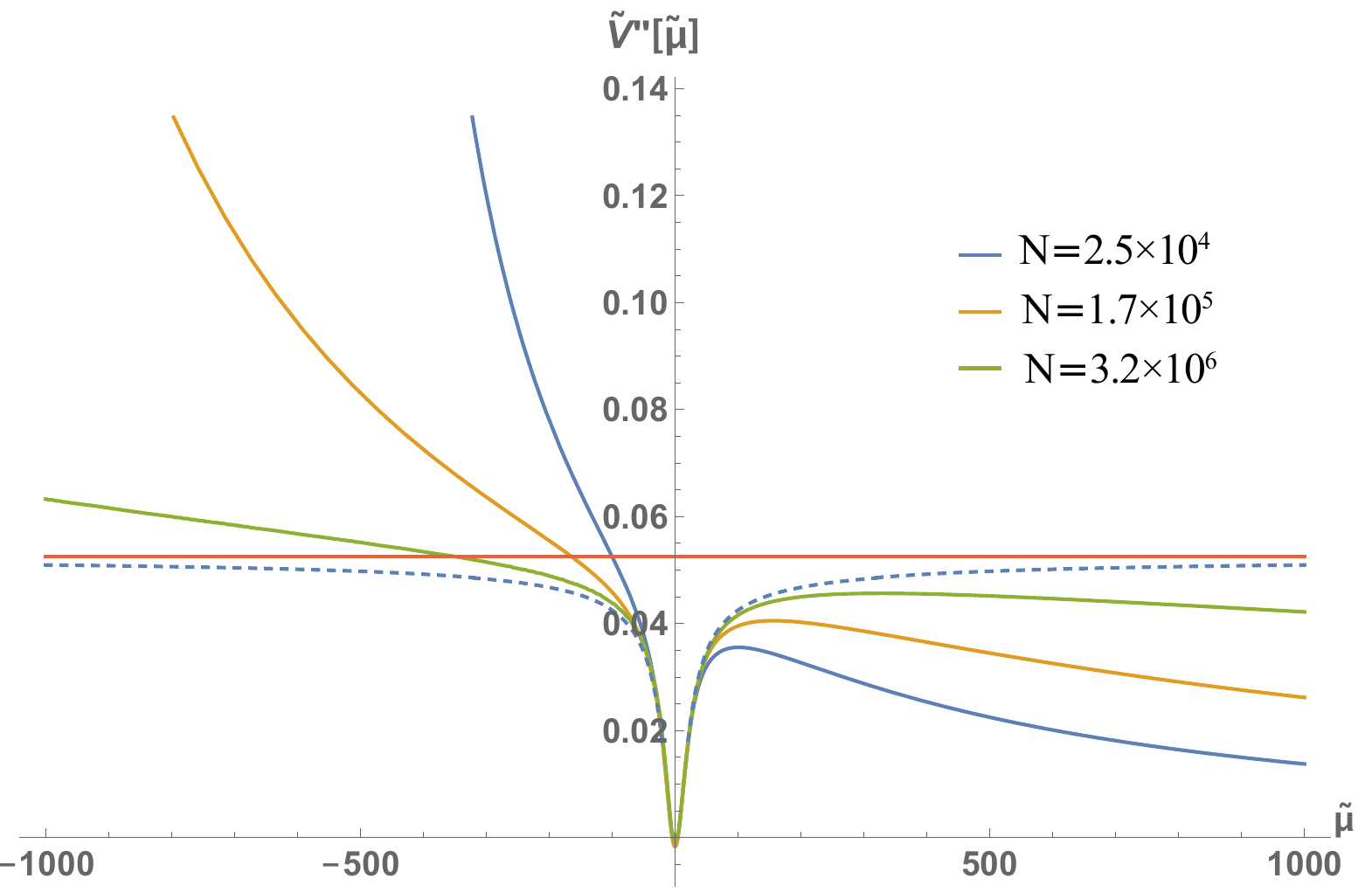}
 \caption{The scaled boundary layer for the second derivative of the $T_{3}$ FP potential $\tilde{V}_N''(\tilde{\mu})$, Eqs. (\ref{flow-LPA-WP-essai-1})  for large values of $N$ in $d=2.6$. The dashed curve  is the global solution $\tilde{V}_{\infty}''(\tilde{\mu})$ of Eq. (\ref{eq:ON-1-1}) at $N=\infty$. The red horizontal line is $y=(-d^2+6d-8)/16$ for $d=2.6$. It coincides with $\tilde{V}_{\infty}''(\tilde{\mu}=\infty)$. }
\label{boundarylayer-1}
\end{figure}

By translating and rescaling by a factor $N^{1/2}$ the position and the width of the boundary layer of the second derivative of the potential $\bar V$, it is possible to obtain a finite limit for this scaled boundary layer when $N\to\infty$. We thus define the scaled variable $\tilde\mu=N^{1/2}(\bar\mu-\bar\mu_0)$ where $\bar\mu_0$ is the location of the boundary layer and the scaled potential by  $\tilde{V}_N(\tilde\mu)=N\left( \bar{V}(N^{-1/2} \tilde\mu +\bar\mu_0)-1/d\right)$. It follows from the definitions above that $\tilde{V}_N''(\tilde\mu)=\bar{V}''(\bar\mu_0+N^{1/2}\tilde\mu)$. We show in Fig. \ref{boundarylayer-1} this scaled boundary layer for different values of $N$ at large $N$ as well as its limit $\tilde{V}_{\infty}''(\tilde{\mu})$ at $N=\infty$.

Notice that a finite difference $\bar\mu-\bar\mu_0$ translates into an infinite $\tilde\mu$ when $N\to\infty$. The matching at $N=\infty$ between the scaled boundary layer and the value of $\bar V''(\bar\mu)$ outside of the layer therefore requires that $\tilde V_\infty''(\infty)=\bar V''(\bar\mu_0^+)=(-d^2+6d-8)/16$ which is the case with the solution for the scaled boundary layer given in the main text, see also Fig. \ref{boundarylayer-1}.

\section{The boundary layer of $\bar U''(\bar\phi)$}

The boundary layer has been derived in the main text in WP version of the RG because it is simpler in this version than in Wetterich version. However, it can also be derived directly in this latter version or, once it is obtained in one version, it can be translated in the other by performing the Legendre transform given in Eq. \eqref{legendre-1}.

\begin{figure}[H]
\includegraphics[scale=0.5]{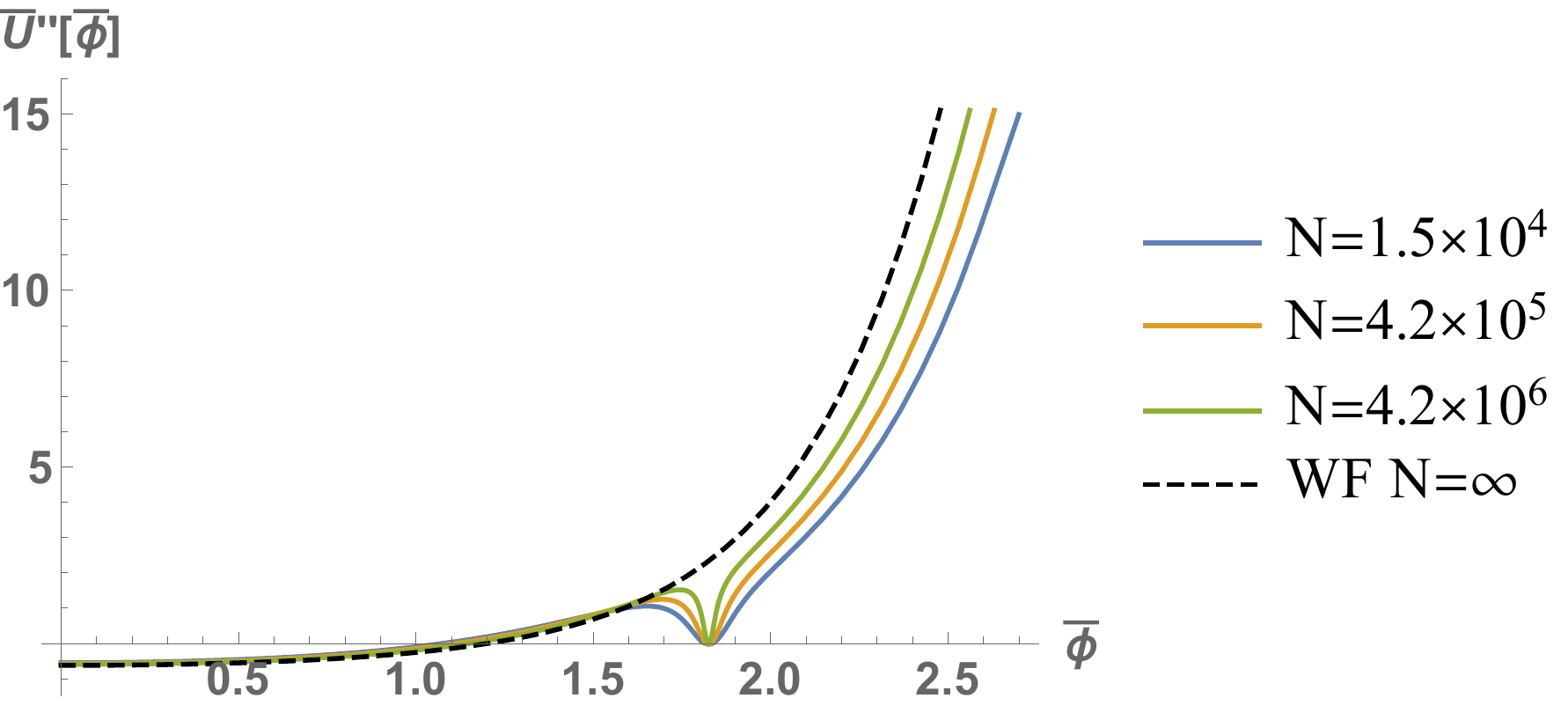}
 \caption{The second derivative of the $T_{3}$ FP potential $\bar{U}''(\bar{\phi})$  in the Wetterich version of the RG, Eq. (\ref{flow-LPA-Ubar-1}), for different values of $N$ in $d=2.6$. }
\label{fig:U''-1}
\end{figure}
We show in Fig. \ref{fig:U''-1} the boundary layer of $\bar{U}''(\bar{\phi})$ for different values of $N$ at large $N$.

\section{Different FP potentials of the BMB line}
\begin{figure}[H]
\includegraphics[scale=0.4]{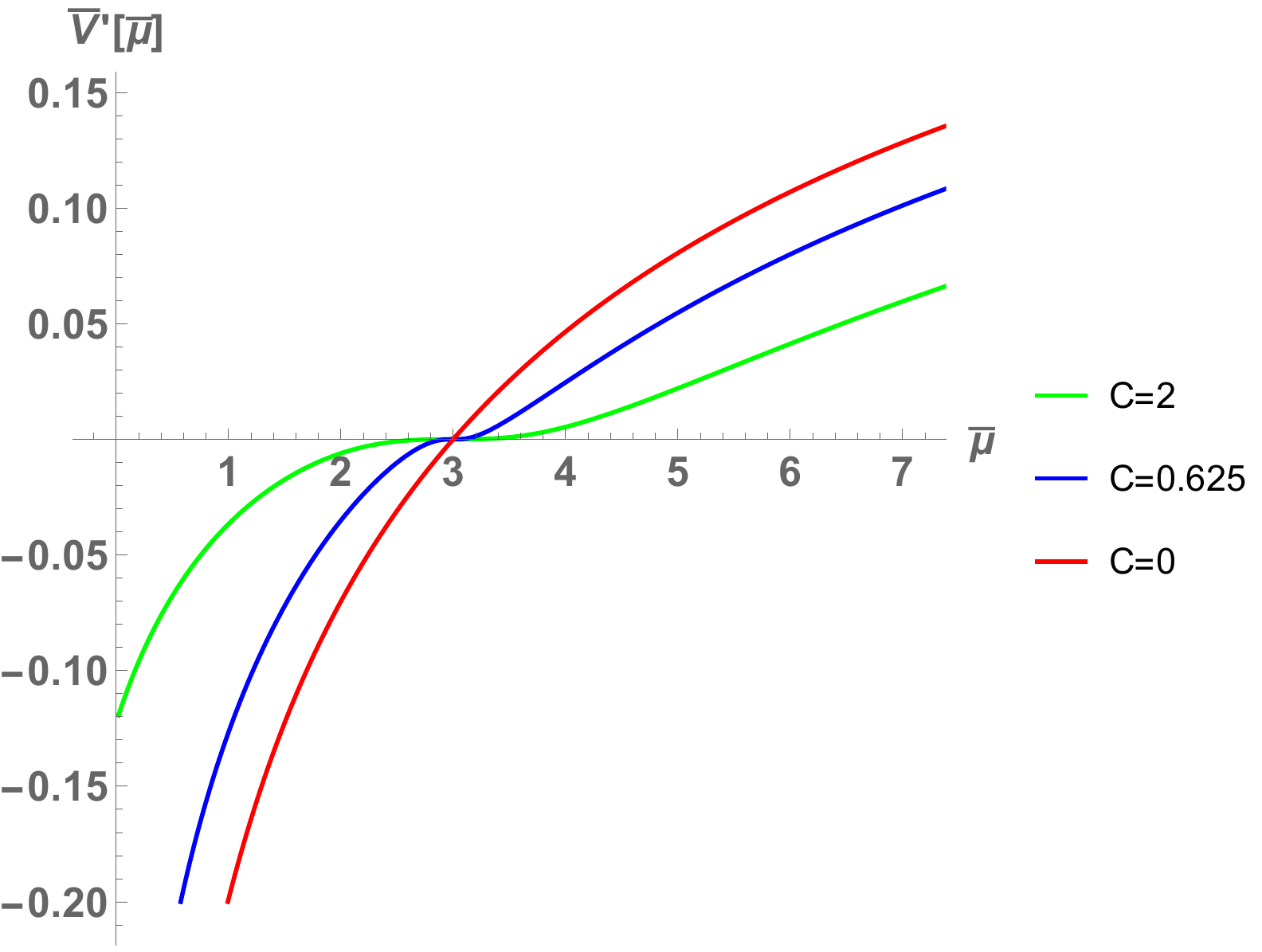}
 \caption{$\bar{V}'(\bar{\mu})$ for different  FPs indexed by the constant $C$ on the BMB line given by Eqs. (\ref{eq:BMB-line-1})  and \eqref{eq:fvprime-1} of the main text.}
\label{fig:BMB-1}
\end{figure}

We show in Fig. \ref{fig:BMB-1} the first derivative of different FP potentials of the BMB line at $N=\infty$ and in $d=8/3$. These FP potentials, implicitly given by the exact expression given in Eqs. \eqref{eq:BMB-line-1} and \eqref{eq:fvprime-1} of the main text, are indexed by the nonnegative constant $C$. The WF FP potential corresponds to $C=0$.

We emphasize that the limit of $\bar V''(\bar\mu_0=3)$, when $C\to 0$ is {\it not} given by the second derivative of the WF FP potential which is however the limit when $C\to 0$  of $\bar V(\bar\mu)$ along the BMB line. This is consistent with what happens at fixed $d<8/3$ when $N\to\infty$ since the limit $d\to8/3$ at fixed $\alpha=\infty$ consists in following the WF FP at $N=\infty$ up to $d=8/3$, the derivatives of which are not the limit of the derivatives of the $T_3$ potential. 

\section{Eigenperturbations at the tetracritical FP}

\begin{figure}[H]
\includegraphics[scale=0.25]{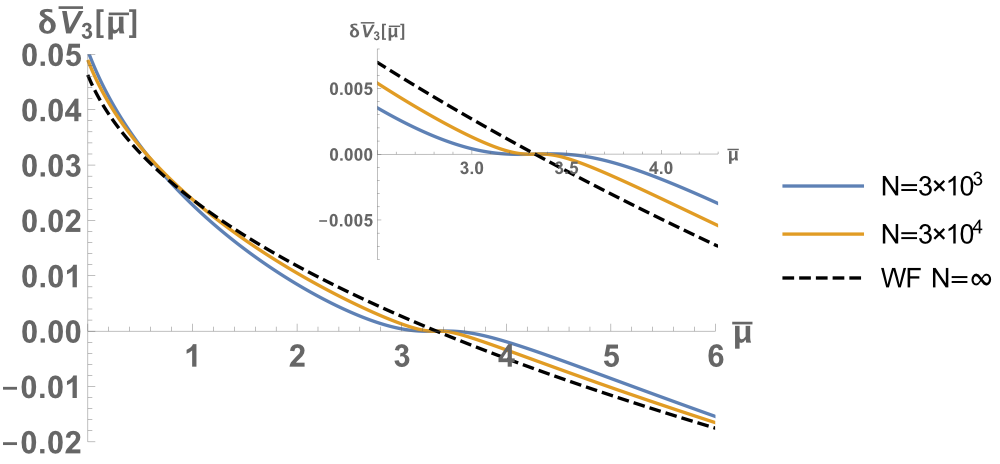}
 \caption{$d=2.6$: Eigenperturbation $\delta \bar{V}_3 (\bar{\mu})$ at the $T_3$ FP corresponding, when $N\to\infty$, to the relevant eigenvalue $\lambda_3=d-2$. } 
\label{fig:eigenperturbation-3-1}
\end{figure} 

We show in Figs. \ref{fig:eigenperturbation-3-1} and \ref{fig:eigenperturbation-1,2-1} the relevant eigenperturbations $\delta \bar V_i$ of the $T_3$ FP in $d=2.6$ for different values of $N$. Whereas $\delta \bar V_3$ tends to the relevant eigenperturbation of the critical WF FP --with eigenvalue $d-2$ which is the inverse of the critical exponent $\nu_{\rm WF}$--, the two others become singular in the $N\to\infty$ limit. This is the reason why they play no role for the critical behavior of the O($N$) model at $N=\infty$.

\begin{figure}[H]
\includegraphics[scale=0.5]{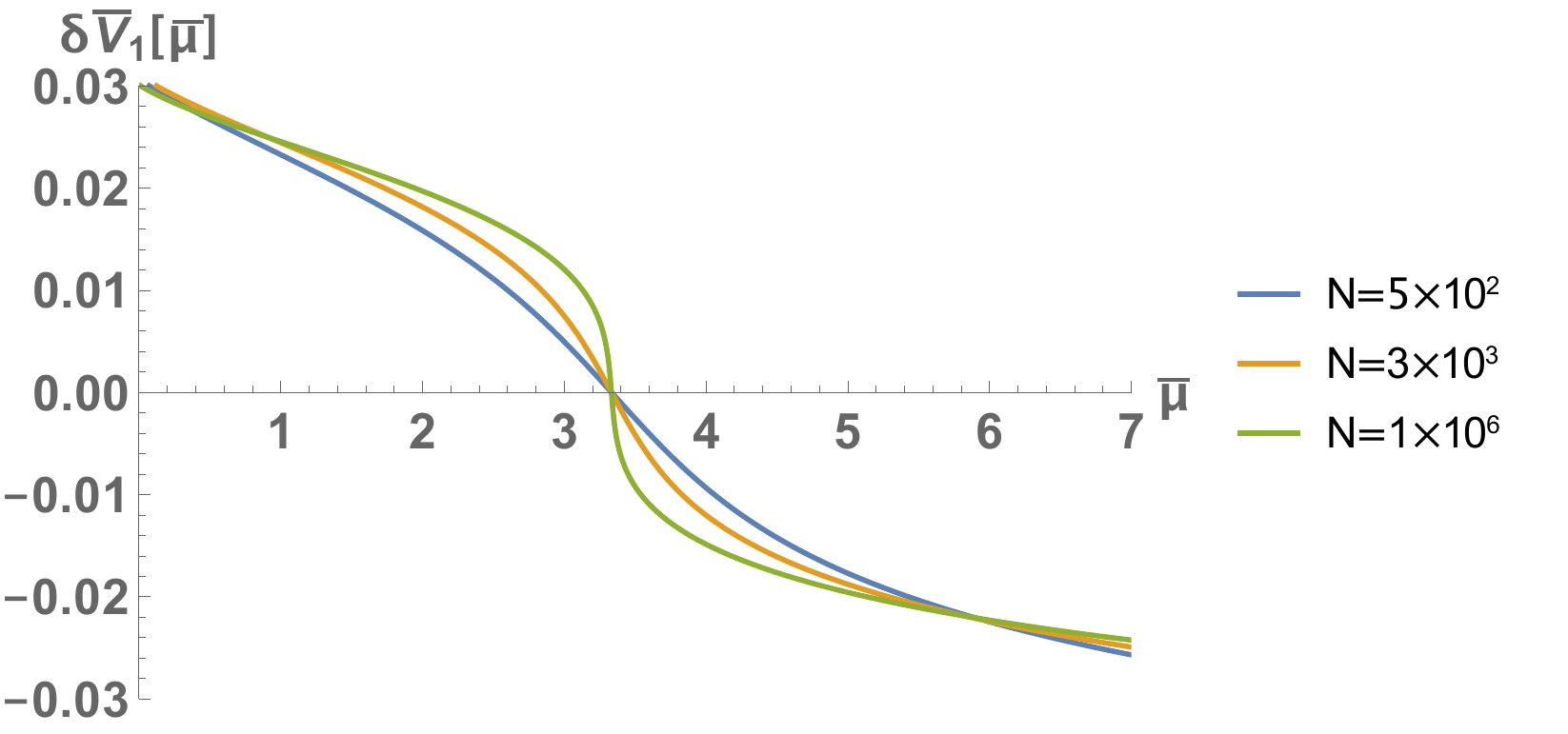}

\ \ \includegraphics[scale=0.5]{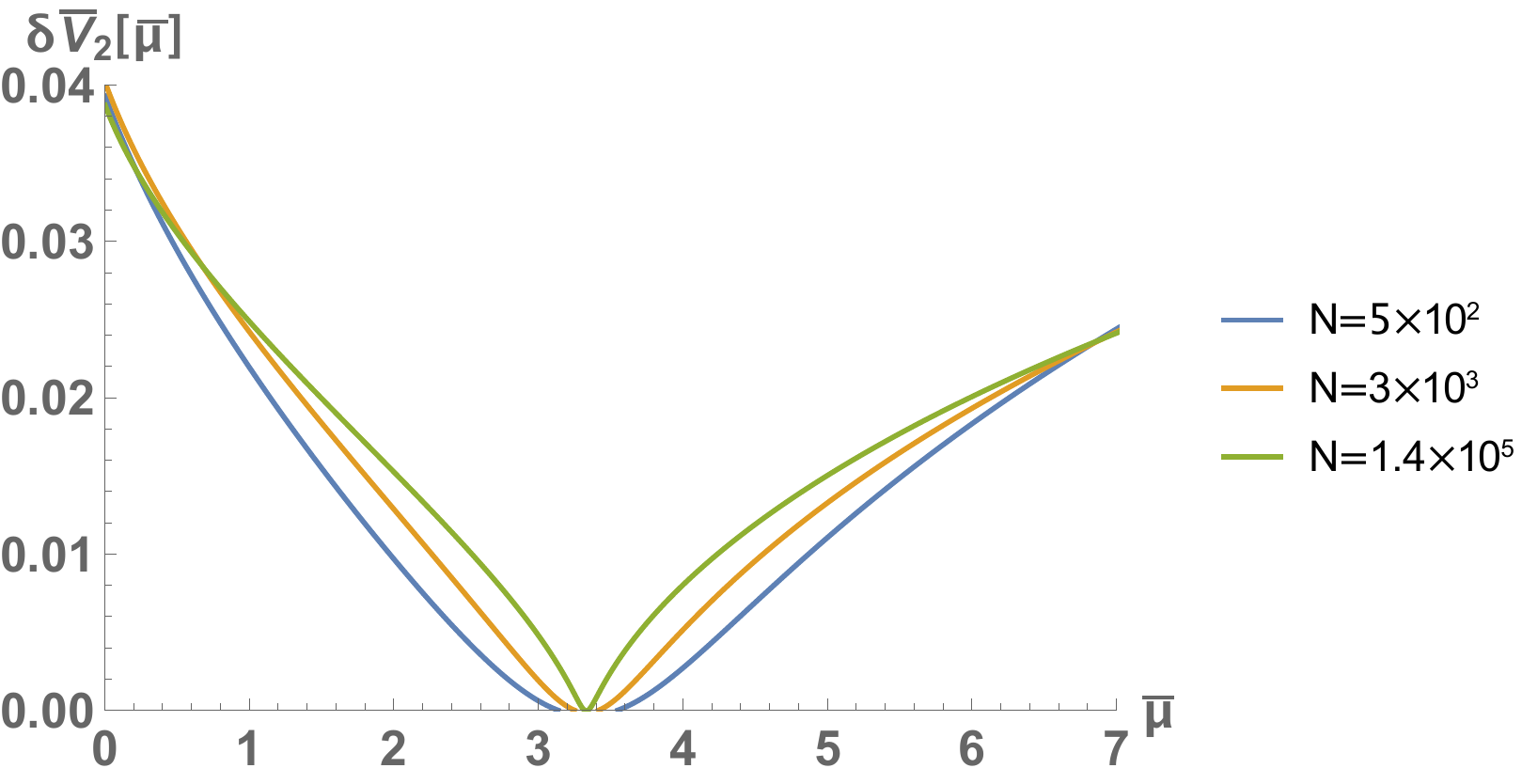}
 \caption{$d=2.6$: Eigenperturbations $\delta \bar{V}_n (\bar{\mu})$ for $n=1,2$ at the $T_3$ FP corresponding respectively to the relevant eigenvalues $\lambda_1\simeq2.00$ and $\lambda_2\simeq1.326$  for different values of $N$. These eigenperturbations tend to singular functions of $\bar\mu$ when $N\to\infty$.  } 
\label{fig:eigenperturbation-1,2-1}
\end{figure}

\begin{figure}[H]
\includegraphics[scale=0.5]{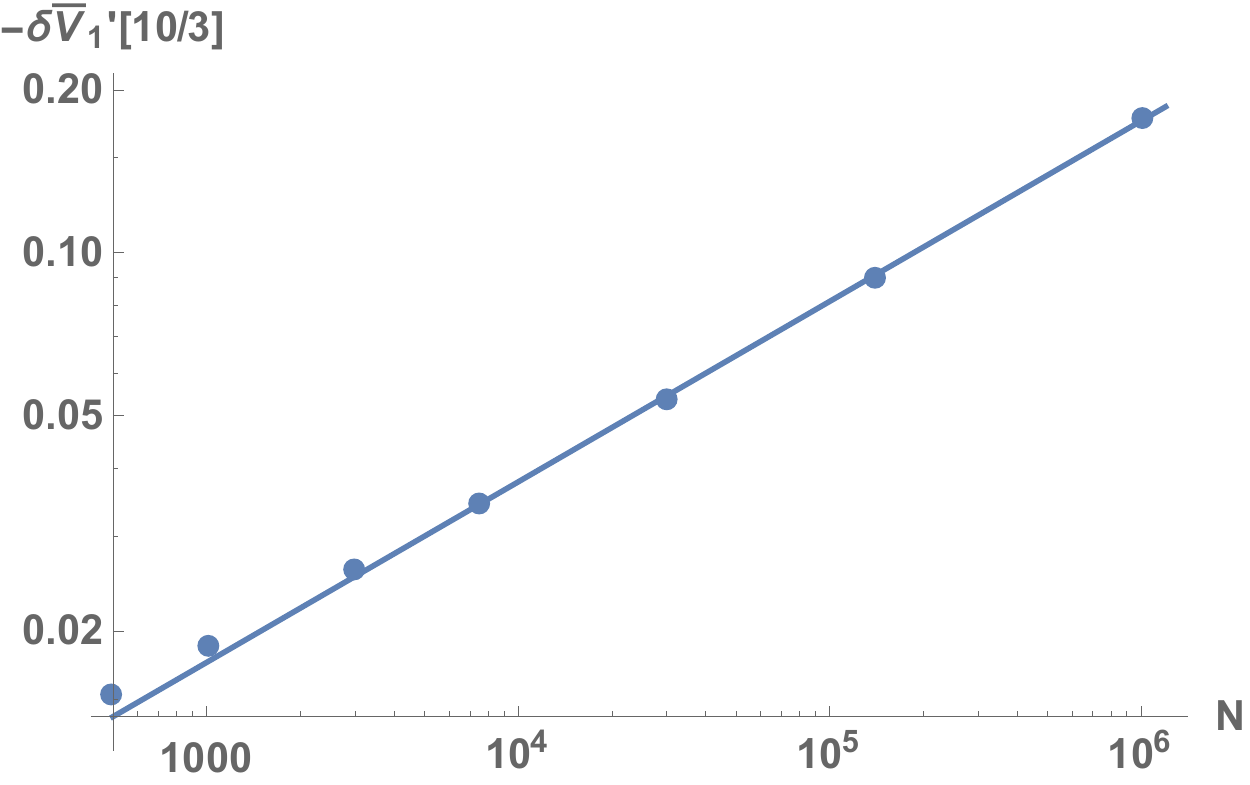}
 \caption{$d=2.6$: Slope of the eigenperturbation $\delta \bar{V}_1(\bar\mu) $ of the $T_3$ FP   at its minimum $\bar{\mu}_0=10/3$ for different values of $N$. The equation of the full line is  $y=0.00175 N^{1/3}$. } 
\label{fig:slope-1}
\end{figure}

We show in Fig. \ref{fig:slope-1} that the slope of $\delta \bar{V}_1(\bar\mu)$ at $\bar\mu_0$ increases as $N^{1/3}$ which proves that this eigenperturbation becomes discontinuous at infinite $N$.

\section{BMB line and the joined limit $\epsilon\to 0$ and $N\to\infty$ at fixed $\alpha=\epsilon N$}

\begin{figure}[H]
\includegraphics[scale=0.63]{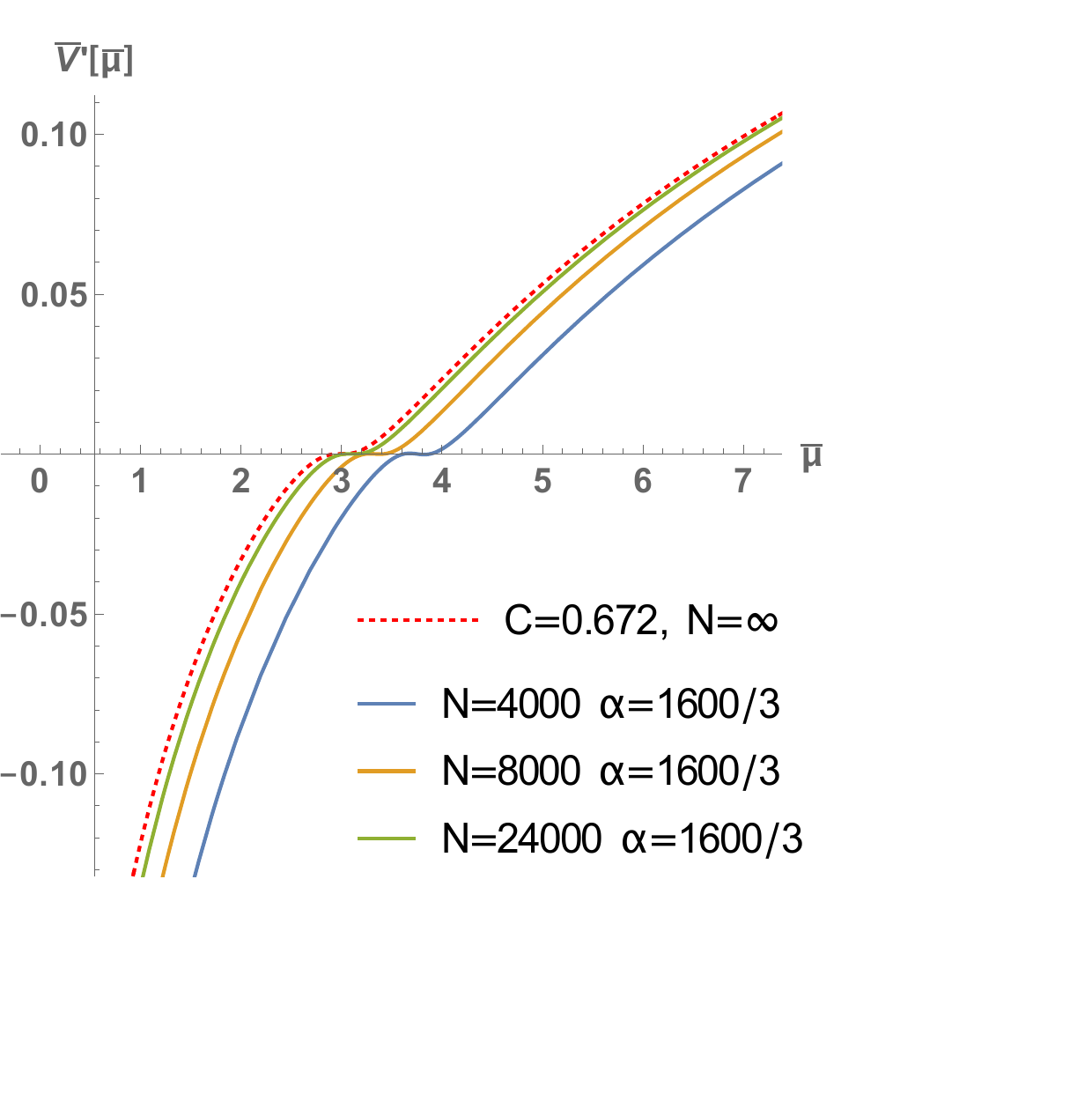}
 \caption{$\bar{V}'(\bar\mu)$ of the $T_3$ FP followed on the hyperbola $d=8/3-\alpha/N$ with fixed $\alpha=1600/3$ for increasing values of $N$. In the double limit $d\to 8/3$ and $N\to\infty$, it converges to the FP potential of the BMB line corresponding to $C=(162 \times 3/1600)^{1/3}\simeq 0.672$. } 
\label{fig:fixedalpha-1}
\end{figure}

When a $T_3$ FP is followed along the hyperbola $d=8/3 - \alpha/N$, $\alpha\ge0$, of the $(d,N)$ plane, its potential converges  when $N\to\infty$ to  the  potential of one of the FPs of the BMB line, see Fig. \ref{fig:fixedalpha-1}. We derive below the relationship $\alpha=162/C^3$ between the parameter $\alpha$ of the hyperbola and the parameter $C$ that indexes the FPs along the BMB line, see Eqs. \eqref{eq:BMB-line-1} and \eqref{eq:fvprime-1} of the main text.

This relationship can be derived as follows. The FP potential of $T_3$ is expanded as 
\begin{equation}
    \bar{V}(\bar\mu)=\sum_{n=0}^{\infty} a_n (\bar\mu -3)^n
\end{equation}
 around the minimum $\bar\mu_0=3$ of the $N=\infty$ potential. Then, the coefficients $a_n$ are expanded as 
 \begin{equation}
   a_n=a_n^{(0)}+N^{-1} a_n^{(1)}+O(N^{-2})
\end{equation} 
in power of $1/N$. At order $O(N^0)$, Eq. (\ref{flow-LPA-WP-essai-1}) yields $a_n^{(0)}=0$ for $n=1,2$ and 3 and recursively determines $a_n^{(0)}$ for $n$ larger than 5 in terms of $a_4^{(0)}$. 

Now, $\sum_{n=0}^{\infty} a_n^{(0)} (\bar\mu -3)^n$ is the expansion of a FP potential of the BMB line. For this potential, $\bar{\mu}_{\pm}$ behaves from Eq. (\ref{eq:BMB-line-1})  as $\bar{\mu}_{\pm}\simeq 3\,\pm\,  2^{4/3}C |\bar{V}'|^{1/3} $ for $C\ne 0$ and $|\bar{V}'|\ll 1$. This implies that $a_4^{(0)}$ is related to $C$ by  $a_4^{(0)}=3/(192 C^3)$. At order $O(N^{-1})$, it can be shown that $a_n^{(1)}$ for all $n$ but $n=4$ can be recursively determined in terms of $a_4^{(0)}$ or $C$, if and only if the condition $\alpha=162/C^3$ is satisfied which proves the relationship between these two parameters.

We show in Fig. \ref{fig:fixedalpha-1} different $T_3$ FP potentials along an hyperbola $d=8/3- \alpha/N$ with increasing values of $N$. 
These FP potentials converge to a potential corresponding to the FP on the BMB line indexed by $C=(162/\alpha)^{1/3}$.

\ 

\newpage


\begin{thebibliography}{1}

\appendix
\bibitem{Pelaez21} M. Pel\'aez, U. Reinosa, J. Serreau, M. Tissier, and N. Wschebor, Rept. Prog. Phys. {\bf 84}, 124202 (2021).


\bibitem{Hoyer21} P. Hoyer, {\it Journey to the Bound States }, Springer (2021).


\bibitem{Guo19} X.-Y. Guo, Y. Heo, and M.F.M. Lutz, Phys. Lett. B
{\bf 791},  86 (2019)

\bibitem{Canet11} L. Canet, H. Chat\'e, B. Delamotte, and N. Wschebor, Phys. Rev. E {\bf 84},
061128 (2011); Phys. Rev. E {\bf 86}, 019904(E) (2012); T. Kloss, L. Canet, B. Delamotte, and N. Wschebor,
Phys. Rev. E {\bf 89}, 022108 (2014).

\bibitem{Gredat}H.-K. Janssen, F. van Wijland, O. Deloubri\`ere,  and U.C. T\"auber, Phys. Rev. E {\bf 70}, 056114 (2004); D. Gredat, H. Chat\'e,  B. Delamotte, and I. Dornic, Phys. Rev. E {\bf 89}, 010102(R) (2014).

\bibitem{tissier08} M. Tissier and G. Tarjus, Phys. Rev. B {\bf 78}, 024204
(2008); 
Phys. Rev. Lett. {\bf 107}, 041601 (2011);
Phys. Rev. B {\bf 85}, 104202 (2012); 
Phys. Rev. B {\bf 85}, 104203 (2012).


\bibitem{Bardeen} W. A. Bardeen, M. Moshe, and M. Bander,
Phys. Rev. Lett. {\bf 52}, 1188 (1984).

\bibitem{Fleming20} C. Fleming, B. Delamotte, and S. Yabunaka, Phys. Rev. D {\bf 102}, 065008 (2020).

\bibitem{BrezinWallace}T. H. Berlin,
and M. Kac, Phys. Rev. {\bf 86}, 821 (1952); H. E. Stanley, Phys. Rev. {\bf 176}, 718 (1968); E. Br\'ezin and D. J. Wallace,
Phys. Rev. B {\bf 7}, 1967 (1973);  S.k. Ma, J. Math. Phys. {\bf 15}, 1866 (1974).

\bibitem{Eyal} G. Eyal, M. Moshe, S. Nishigaki, and J. Zinn-Justin, Nucl. Phys. B {\bf 470} [FS], 369 (1996).

\bibitem{Zinn-Justin} J. Zinn-Justin, {\it Quantum field theory and critical phenomena Fourth Edition}, Oxford University Press, (2002).






\bibitem{Itzykson}
 C. Itzykson and J. Drouffe, {\it Statistical Field Theory: Volume 1, From Brownian Motion to Renormalization and Lattice
Gauge Theory}, Cambridge University Press, (1989).

\bibitem{defenu2020} N. Defenu and A. Codello, ArXiv:2005.10827.

\bibitem{Comellas}
J. Comellas and A. Travesset, Nucl. Phys. B {\bf 498}, 539 (1997).

\bibitem{Wilson} K. G. Wilson and J. Kogut, Phys. Rep., Phys. Lett.,{\bf 12C} (1974) 75.

\bibitem{PhysRevB.4.3174} K. G. Wilson, Phys. Rev. B {\bf 4}, 3174 (1971).

\bibitem{wetterich91} C. Wetterich, Nucl. Phys. B {\bf 352}, 529 (1991).
\bibitem{wetterich93b} C. Wetterich, Phys. Lett. B {\bf 301}, 90 (1993).
\bibitem{Ellwanger} U. Ellwanger, Z. Phys. C. Part. Fields {\bf 58}, 619 (1993).
\bibitem{Morris94} T. R. Morris, Int. J. Mod. Phys. A {\bf 09}, 2411 (1994).

\bibitem{Berges}J. Berges, N. Tetradis and C. Wetterich, Phys. Rept. {\bf 363}, 223 (2002)

\bibitem{Delamotte-review} B. Delamotte, D. Mouhanna, and M. Tissier, Phys. Rev. B {\bf 69}, 134413 (2004).

\bibitem{Delamotte-lect-notes} B. Delamotte, “{\it An Introduction to the nonperturbative renormalization group}”, Lect. Notes Phys. {\bf 852}, 49 (2012).

\bibitem{canet03} L. Canet, B. Delamotte, D. Mouhanna, and J. Vidal,
Phys. Rev. B {\bf 68}, 064421 (2003).

\bibitem{canet05} L. Canet, Phys. Rev. B {\bf 71}, 012418 (2005).

\bibitem{kloss14} T. Kloss, L. Canet, B. Delamotte, and N. Wschebor,
Phys. Rev. E {\bf 89}, 022108 (2014); L. Canet, H. Chat\'e, B. Delamotte, and N. Wschebor, Phys. Rev. E {\bf 84},
061128 (2011); Phys. Rev. E {\bf 86}, 019904(E) (2012).

\bibitem{delamotte04} B. Delamotte and L. Canet, Condensed Matter Phys. {\bf 8},  163 (2005).

\bibitem{benitez08} F. Benitez, R. M\'endez-Galain, and N. Wschebor, Phys.
Rev. B {\bf 77}, 024431 (2008).

\bibitem{canet04} L. Canet, B. Delamotte, O. Deloubri\`ere, and N. Wschebor, Phys. Rev. Lett. {\bf 92}, 195703 (2004); 
L. Canet, H. Chat\'e, and B. Delamotte, Phys. Rev. Lett. {\bf 92}, 255703
(2004); 
L. Canet, B. Delamotte, D. Mouhanna, and J. Vidal, Phys. Rev. D {\bf 67}, 065004 (2003); 
L. Canet, H. Chat\'e, B. Delamotte, I. Dornic, and M. A. Mu\~noz,
Phys. Rev. Lett. {\bf 95}, 100601 (2005).

\bibitem{canet16} L. Canet, B. Delamotte, and N. Wschebor, Phys. Rev.
E {\bf 93}, 063101 (2016).

\bibitem{leonard15} F. L\'eonard and B. Delamotte, Phys. Rev. Lett. {\bf 115}, 200601 (2015).

\bibitem{balog19}Balog, Ivan and Chat\'e, Hugues and Delamotte, Bertrand and Marohni\ifmmode \acute{c}\else \'{c}\fi{}, Maroje and Wschebor, Nicol\'as, Phys. Rev. Lett. {\bf 123}, 240604 (2019)

\bibitem{Tetradis-Litim}N. Tetradis and D. F. Litim, Nucl. Phys.
B {\bf 464}, 492 (1996).

\bibitem{dattanasio} M. D'Attanasio and T. R. Morris, Phys. Lett. B {\bf 409}, 363 (1997).

\bibitem{Kubyshin}Y. Kubyshin, R. Neves, and R. Potting, Int. J. Mod. Phys. A {\bf 16}, 2065 (2001). 

\bibitem{Katis-Tetradis}A. Katsis and N. Tetradis, Phys. Lett. B {\bf 780}, 491-494 (2018). 


\bibitem{David} F. David, D. A. Kessler, and H. Neuberger, Phys. Rev. Lett. {\bf 53}, 2071 (1984).
\bibitem{Omid} H. Omid, G. W. Semenoff, and L. C.R. Wijewardhana, Phys. Rev. D {\bf 94}, 125017 (2016).
\bibitem{Mati2017} D. F. Litim, E. Marchais, and P. Mati, Phys. Rev. D {\bf 95}, 125006 (2017).

\bibitem{Morris}T. R. Morris,    J. High Energy Phys., {\bf 07}, 027 (2005).

\bibitem{Polchinski}J. Polchinski, Nucl. Phys. B {\bf 231} (1984) 269.

\bibitem{Hasenfratz}A. Hasenfratz and P. Hasenfratz, Nucl. Phys. B, {\bf 270}, 687  (1986).

\bibitem{Holmes}M. H. Holmes, {\it Introduction to perturbation methods Second edition}, Springer (2012).

\bibitem{footnote1}Note that the first term in Eq. (\ref{eq:ON-1-1}) comes from the last term in Eq. (\ref{flow-LPA-Ubar-1}) or Eq. (\ref{flow-LPA-WP-essai-1}), which is formally proportional to $N^{-1}$ and neglected in the usual large-$N$ analysis. However this term is indispensable to describe the boundary layer of $\bar U''(\bar\phi)$ or $\bar V''(\bar\mu)$.

\bibitem{Pisarski}R. D. Pisarski, Phys. Rev. Lett. {\bf 48}, 574 (1982). 
\bibitem{Osborn} H. Osborn and A. Stergiou, J. High Energy Phys. {\bf 5} 51 (2018).



\bibitem{Yabunaka-Delamotte}S. Yabunaka and B. Delamotte, Phys. Rev. Lett. {\bf 119}, 191602 (2017); Phys. Rev. Lett. {\bf 121}, 231601 (2018).

\bibitem{Yabunaka-PRE2022} S. Yabunaka, C. Fleming, and B. Delamotte,
Phys. Rev. E {\bf 106}, 054105 (2022).

\end{thebibliography}
\end{document}